\documentclass[12pt,endfloat]{article}
\usepackage{varioref}
\usepackage{graphicx}
\usepackage[T1]{fontenc}
\usepackage{epsfig}
\usepackage{subfigure}
\usepackage{cite}
\usepackage[labelsep=space]{caption}
\usepackage{amsmath}
\usepackage{amsfonts}
\usepackage{stmaryrd}
\usepackage{psfrag}
\usepackage{esint}
\usepackage[a4paper]{geometry}
\usepackage{setspace}
\usepackage{color}
\usepackage[utf8]{inputenc}
\usepackage{authblk}
\usepackage{tabularx}\newcolumntype{Y}{>{\centering\arraybackslash}X}

\begin{document}

\title{Contactless optical spinning tweezers with tunable rotation frequency}
\author[1,2]{N. Hameed}
\author[1,3]{T. Zeghdoudi}
\author[1]{B. Guichardaz}
\author[3]{A. Mezeghrane}
\author[1]{M. Suarez}
\author[1]{N. Courjal}
\author[1]{M.-P. Bernal}
\author[3]{A. Belkhir}
\author[1,*]{F. I. Baida}

\affil[1]{Institut FEMTO-ST, UMR 6174 CNRS, D\'epartement d'Optique P. M. Duffieux, Universit\'{e} Bourgogne Franche--Comt\'{e}, 25030 Besan\c{c}on Cedex, France}
\affil[2]{Al Muthanna University, College of Science,Department of Physics, Al Muthanna, Iraq}
\affil[3]{Laboratoire de Physique et Chimie Quantique, Universit\'{e} Mouloud Mammeri, Tizi-Ouzou, Algeria}
\affil[*]{Corresponding author: fbaida@univ-fcomte.fr}

\maketitle



\begin{abstract}
Advances in optical trapping design principles have led to tremendous progress in manipulating nanoparticles (NPs) with diverse functionalities in different environments using bulky systems. However, efficient control and manipulation of NPs in harsh environments require a careful design of contactless optical tweezers. \textcolor{black}{Here, we propose a simple design of a fibered optical probe allowing the trapping of dielectric NP as well as a transfer of the angular momentum of light to the NP inducing its mechanical rotation}. A polarization conversion from \textcolor{black}{linearly-polarized incident guided to circularly transmitted beam} is provoked geometrically by breaking the cylindrical symmetry of a coaxial nano-aperture that is engraved at the apex of a tapered metal coated optical fiber. Numerical simulations show that this simple geometry tip allows powerful light transmission \textcolor{black}{together} with efficient polarization conversion. This guarantees very stable trapping \textcolor{black}{of quasi spherical NPs} in a non-contact regime as well as potentially very tunable and reversible rotation frequencies in both directions (up to 45~Hz in water and 5.3~MHz in air for 10~mW injected power in the fiber). This type of fiber probe opens the way to a new generation of miniaturized tools for total manipulation (trapping, sorting, spinning) of NPs.
\end{abstract}
\section{Introduction}
Since its discovery by Ashkin in 1970 \cite{ashkin:prl70}, optical tweezers have became a tool of choice for physicists, chemists and biologists in the manipulation of nanoparticles (NPs). The advent of metamaterials in the field of optics has boosted research in this direction with the aim of miniaturizing these systems using them to explore \textcolor{black}{physical and chemical properties in the field of the infinitely small scale }(chemical bonds between molecules and even atomic bonds). At these scales, the tools must be adapted in size, so that they induce the minimum disturbance on the quantities to be measured. It is therefore privileged to set up standalone systems allowing both the trapping of NPs and their translation and rotation.
To this end, metamaterials, which have been extensively studied over the past decades, can be of great help. Indeed, they have shown their ability to address a wide \textcolor{black}{variety} of applications ranging from detection \cite{yoo:nanophot19}, beam shaping and polarization \cite{li:nl13,wolf:nc15,MohammadiEstakhri2016,lyer:ieeetap20} to optical trapping through the control of light confinement via changes in the phase and group velocities of the waves passing through them \cite{hong:NN20}. This confinement is essential to the trapping itself and will allow the translation movement. The rotation, as for it, is generally induced by a transfer towards the NP of the angular momentum of the light. This last one requires the control of the \textcolor{black}{light polarization} at nanometric scales. 
Consequently, the design of \textcolor{black}{standalone miniaturized optical tweezers lacks only one element which consists in the miniaturization of the wave-plates (quarter- and half-wave plates)}. 
However, optical \textcolor{black}{rotation} of NPs was already demonstrated since 1997 \cite{simpson:ol97} and has seen considerable growth both theoretically and experimentally over the past 20 years \cite{friese:nat98,friese:apl01,dholakia:pw02,leach:lc06,perkins:lpr09,tong:nl10,balushi:ana15,keller:mm18}. Most of studies addressed applications in biology \cite{perkins:lpr09,keller:mm18,zhang:jbio19,lee:sr21}. Recently, Lehmuskero \emph{et al.} \cite{lehmuskero:nl13} demonstrated \textcolor{black}{gold} NP spinning within a circularly polarized focused beam. They measured \textcolor{black}{spinning frequencies of several kHz for NP rotating around its center of mass}. Nevertheless, they only had a 2D trapping (in the transverse plane) meaning the NP is pushed against a glass slide to restrict its motion along \textcolor{black}{the beam propagation direction}. Some authors \cite{gieseler:prl12,reimann:prl18,ahn:prl18,monteiro:pra18} recently showed optical trapping in vacuum accompanied with high spinning motion \textcolor{black}{around the beam axis} with frequency beyond 1GHz but their schemes involve bulky systems composed of propagating beams, conventional optical elements such as lenses, polarizers, wave-plates and/or diaphragms which makes the system very cumbersome. \\

\begin{figure}
\centering
\includegraphics[width=8.5cm]{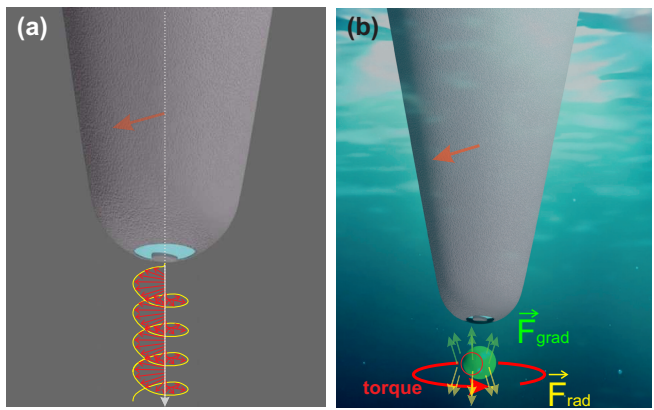}
\caption{\textbf{Schematic of the proposed Optical Spinning-Trapping Tweezers (OSTT): (a)} Principle of the polarization conversion. \textbf{(b):} Artistic view of the simultaneous trapping and spinning of a NP.}
\label{idee}
\end{figure}

In this study, our fibered Optical Spinning-Trapping Tweezers (OSTT) combines simultaneously a contactless trapping and a linear-to-circular polarization conversion (see Fig. \ref{idee}) \textcolor{black}{allowing the NP translation simultaneously with its rotation. As it will be demonstrate in the following, the designed OSTT is able to induce a spinning of a quasi-spherical NP around its center of mass even if it is made of isotropic loss-less dielectric material. In fact, a real NP is never perfectly spherical nor homogeneous at the nanometer scale. The FDTD method, based on the discretization of the space into small parallelepipeds, inevitably and naturally leads to an approximate geometry of the sphere (staircase effect). This leads directly to the appearance of a dissymmetry due the mesh practiced in FDTD (see Appendix A) that has already been numerically exploited to demonstrate the excitation of BICs (Bound states In the Continuum) and/or (SPMs) symmetry protected modes \cite{hoblos:ol20,hoblos:oc20}. In addition,} the inhomogeneity of the transmitted field, due the presence of evanescent components, \textcolor{black}{helps the spin angular momentum transfer from light to the NP}. When operating in vacuum, this could achieve spinning motion of trapped NPs \textcolor{black}{around their center} with relatively high frequency. The control of the NP rotation frequency could easily be performed by simply changing the polarization direction of the guided wave inside the \textcolor{black}{monomode} optical fiber \textcolor{black}{through the use of} a conventional polarization controller. \\

\section{Proposed structure and geometry optimization}
\begin{figure}[h!]
\centering
\includegraphics[width=8.5cm]{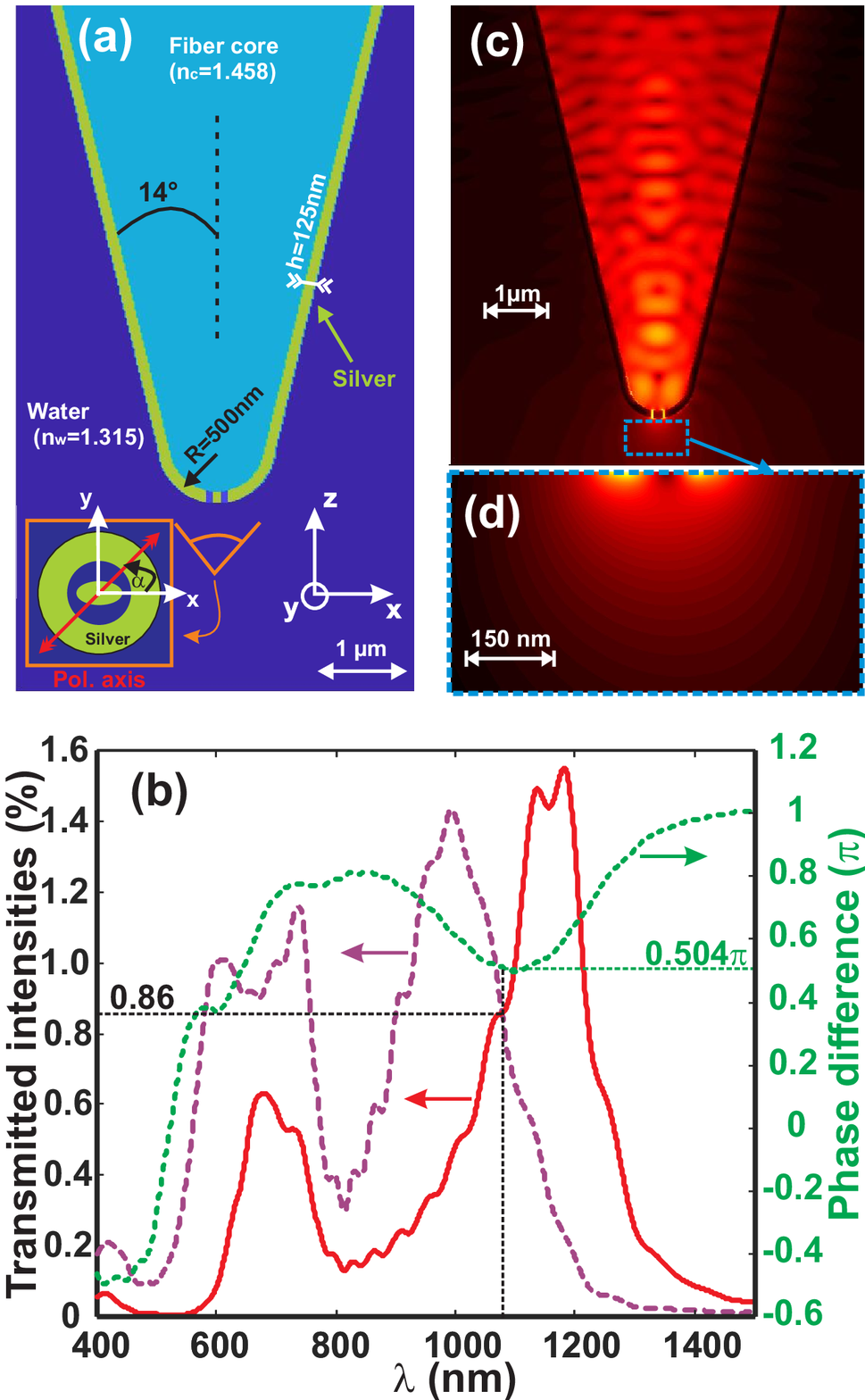}
\caption{\textbf{(a) Schematic of the modeled OSTT.} All geometric parameters and electromagnetic constants are given in the figure. The bottom left inset shows the polarization direction of the guided mode which is oriented at $\alpha $~=~45$^\circ$ from the ellipse axes needed to get circularly polarized transmitted beam. \textbf{(b) Transmission spectra} of the two electric-field components (x in red solid line and y in purple dashed line) associated with the zero-order diffracted light for a silver coating thickness of $h $~=~125~nm. The intersection of these two spectra corresponds to the operating point (quarter-wave plate) because the phase difference between the two transverse components of the electric field is well equal to $\pi/$2 as shown by the dotted green curve. \textbf{(c) Electric field distribution} (fifth root of the intensity) in a vertical plane at the operation wavelength. \textbf{(d) Zoom-in} on the area in front of  {the OSTT apex of} Fig. (c) showing the electric field gradient  {necessary} for trapping.}
\label{schemasonde}
\end{figure}

Our structure is based on an individual coaxial aperture engraved on the apex of a metal (silver for instance) coated tapered single-mode optical fiber. The geometry of the aperture was optimized to exhibit an efficient transmission coefficient and a robust polarization conversion versus the fabrication uncertainties. To this end, we made extensive Finite Difference Time Domain (FDTD) numerical simulations using home-made codes ({See Appendix A}) to adapt the metal thickness to the desired properties. The polarization conversion is obtained by light transmission through a geometrical anisotropy of the coaxial aperture \cite{dahdah:ieee12} \textcolor{black}{(elliptical inner part instead of circular one)}. The origin of this anisotropy is linked to the excitation of two orthogonally-polarized (TE$_{11}$) guided modes inside the metal thickness having plasmonic/propagative hybrid character \cite{baida:prb06} with different effective indices. This anisotropy was described in details in refs. \cite{baida:prb11,dahdah:ieee12,boutria:prb12}. It could \textcolor{black}{be exploited to induce the rotation of the NP provided that it allows the necessary conversion of the linear polarization (in the fiber) into circular polarization (at the tip output)}. Nevertheless, a stable trapping is simultaneously necessary to fix the position of the NP in front of the tip. Therefore, electromagnetic confinement within the aperture is essential to induce a sufficient gradient force capable of compensating for the radiation pressure induced by the light flow (funnel effect) through the aperture. 
The geometry of the structure has been optimized \textcolor{black}{by adapted calculations so that they are less costly in terms of} computing time and memory space. \textcolor{black}{This consisted of two steps starting from the consideration of a 2D grating of coaxial apertures and moving towards a single aperture (please refer to the Appendix B). The latter is then transferred to the apex of a SNOM (Scanning Near-field Optical Microscope) probe that consists on a metal-coated tapered monomode optical fiber}. The fiber core index is $n_c $~=~1.458 and the cladding one is $n_g $~=~1.453. The core diameter is set to $D_c $~=~4.2 $\mu$m as for a SMF-28 or X1060 fibers from Thorlabs. The cone angle of the tip has a typical value\cite{bhushan:book06} of 28$^\circ$. \textcolor{black}{In the FDTD simulations, the fundamental guided mode of the fiber is injected into the upper part of the probe (see Fig. \ref{schemasonde}a) at more than 5$\mu$m from the apex. The geometrical parameters of the elliptical coaxial aperture are given as in Appendix B. The thickness of the metallic layer is set to $h=125~nm$ allowing the aperture to behave as a quarter-wave plate (the two transmitted transverse components of the electric field have the same amplitude and are phase shifted by $\pi/2$ as presented in Fig. \ref{schemasonde}b). In this figure, the transmitted intensities are calculated by normalizing the transmitted energy associated with each electric field component to the total energy injected into the fiber. The operation wavelength is then around $\lambda=1080~nm$.} The temporal variation of \textcolor{black}{the guided mode} electric field (amplitude and direction) in a transverse plane intersecting the fiber at $z $~=~4~$\mu$m from its apex is shown in the video \textbf{"Visualisation1.avi"} when the polarization of the guided mode is supposed to be directed at $\alpha $~=~45$^\circ$ with respect to the axes of the ellipse (see inset of Fig. \ref{schemasonde}a).
\begin{figure}
\centering
\includegraphics[width=16cm]{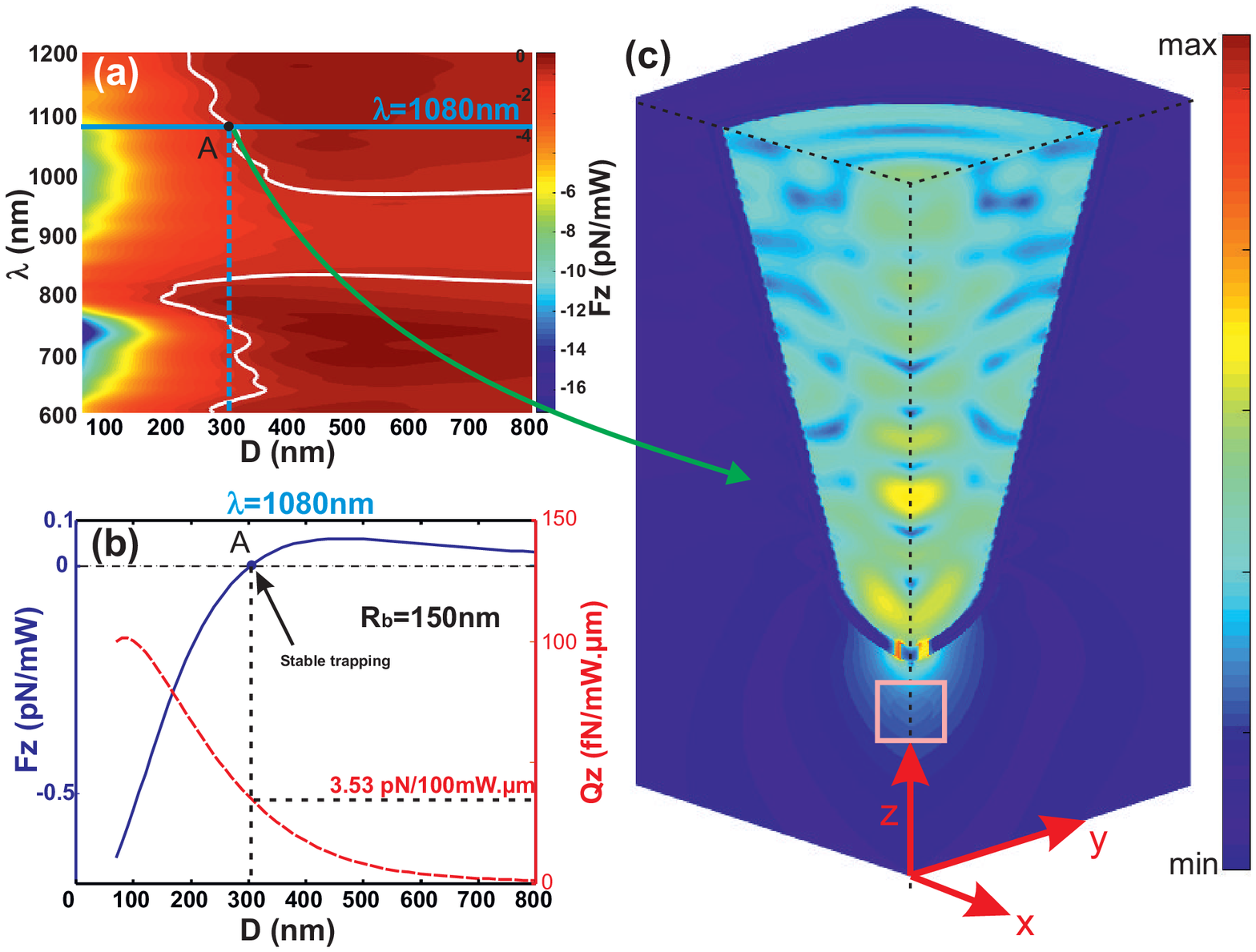}
\caption{\textbf{Trapping and Spinning of a {$R_b=150$}~nm -radius NP. (a) Vertical force} $F_z$ exerted on a 150~nm radius latex NP as a function of the injected wavelength and the OSTT-to-NP distance $D$ . The white line denotes the NP position for which the optical force vanishes. The blue horizontal line corresponds to the quarter-wave operation wavelength ($\lambda $~=~1080~nm). \textbf{(b) Magnitude of the z-component of the force and the torque} for the same NP ($R_b $~=~150~nm, $n_b $~=~1.5) as a function of $D$ at $\lambda $~=~1080~nm. Point A ($D $~=~305~nm) corresponds to stable trapping  {($F_z$ vanishes) while} the z-component of the torque does not { cancel}. \textbf{(c) Electric intensity distributions} ({fifth root}) in three perpendicular faces of a rectangular parallelepiped corresponding to the operation point A (trapping + spinning). The OSTT-axis coincides with the intersection of the two vertical planes. \textcolor{black}{If we look closely inside the pink square, we could distinguish the presence of the trapped NP}. }
\label{Rb150}\rule{\textwidth}{0.4pt}
\end{figure}
As it can be seen from Fig. \ref{schemasonde}b, the intersection of the two spectra should correspond to the required operating wavelength for which the phase difference $\Delta\phi=Arg(E_x/E_y)$ between the two transverse components ($E_x$ in red solid line and $E_y$ in purple dashed line) is equal to $\pi/$2. The variations of $\Delta\phi$ are \textcolor{black}{also} plotted in green dotted line on the same figure. One can clearly see the latter condition to be fulfilled exactly for the wavelength value of $\lambda $~=~1080~nm corresponding to the abscissa of the intersection point. In addition, the positive value of $\Delta\phi$ means that the $E_x$ is in phase advance with respect to $E_y$. This indicates that the emitted wave \textcolor{black}{by the aperture} is circularly right polarized. The temporal evolution of the transmitted electric field (amplitude and direction) by the OSTT in a transverse plane located at $z $~=~350~nm in front of its apex is shown in the video \textbf{"Visualization2.avi"}. A light  angular momentum transfer from photons to the NP will then induce its rotation in the same direction that corresponds to a positive value of $Q_z$ i.e. the z-component of the torque (see the coordinate system of Fig. \ref{schemasonde}a). Note that, at the operation wavelength, the transmission value is around 1.7$\%\approx$2$\times$ 0.86$\%$ (ratio of the transmitted total energy to the injected energy into the fiber), a fairly efficient transmission compared to a simple cylindrical or rectangular aperture \cite{baida:prb06}. Let us notice that the small oscillations appearing on the spectra of Fig. \ref{schemasonde}b have a real physical meaning: they result from interferences between the surface plasmon wave propagating along with the metal coating (\textcolor{black}{probe} external walls) and the transmitted guided field by the aperture itself \cite{baida:oc03}. Figure \ref{schemasonde}c shows the distribution of the electric field (fifth root of the intensity) at the operation wavelength $\lambda $~=~1080~nm in a vertical plane parallel to $xz$ containing the probe axis. Below we present (Fig. \ref{schemasonde}d) an enlargement of the area in front of the OSTT apex showing the large confinement of the electric field i.e. the high gradient needed to get efficient optical trapping.
\section{Optical forces and torques}
To quantitatively evaluate the optical force $\vec{F}$ and the torque $\vec{Q}$ induced by the electromagnetic field transmitted by the probe on a given NP, we use the classical definitions given by:
\begin {equation}
\vec{F} =\oiint_s \stackrel{\leftrightarrow}{T} \cdot \vec ds  \hspace{2cm} \vec{Q} =-\oiint_s \stackrel{\leftrightarrow}{T} \wedge \hspace{.1cm}\vec{r} \cdot\vec ds\label{force1}
\end {equation}
Where, $\stackrel{\leftrightarrow}{T}$ is the well-known Maxwell stress tensor with elements defined by:
\begin {equation}
 {T}_{ij}=\varepsilon_0(E_i\cdot E_j-\frac {1}{2} \Delta_{ij}  E^2)+{\mu_0}(H_i\cdot H_j-\frac {1}{2} \Delta_{ij} H^2)
\end{equation}
Equations \ref{force1} are valid in harmonic regime ($e^{-i\omega t}$).  {Due to the conservative character of the optical forces and using the divergence theorem, the integration domains in both integrals consist of any surface surrounding the NP. Practically, we consider a cube that encompasses the NP and whose faces remain in the same medium (here water).
Figure \ref{Rb150}a shows the evolution of the vertical component $F_z$ of the optical force applied on a NP made in latex ($n $~=~1.5) with a radius $R_b $~=~150~nm as a function of the wavelength and the \textcolor{black}{distance $D$ between the OSTT and the nearest point of the NP}. The NP is supposed to move along the OSTT-axis so that, due to the symmetry of the configuration, only the vertical components of the force and of the torque are predominant. The white line on Fig. \ref{Rb150}a corresponds to the positions where $F_z$ vanishes meaning a possible trapping of the NP. Nonetheless, to be stable, the trapping must corresponds to a potential well meaning a $F_z$ variation from negative to positive values when the distance $D$ increases (see the $xyz$ frame on Fig. \ref{schemasonde}a). As can be seen in Fig. \ref{Rb150}a, this stable trapping condition is well met at the $A$ point corresponding to $D $~=~305~nm and the operating wavelength of the quarter-wave plate ($\lambda $~=~1080~nm).
\begin{figure}
\centering
\includegraphics[width=16cm]{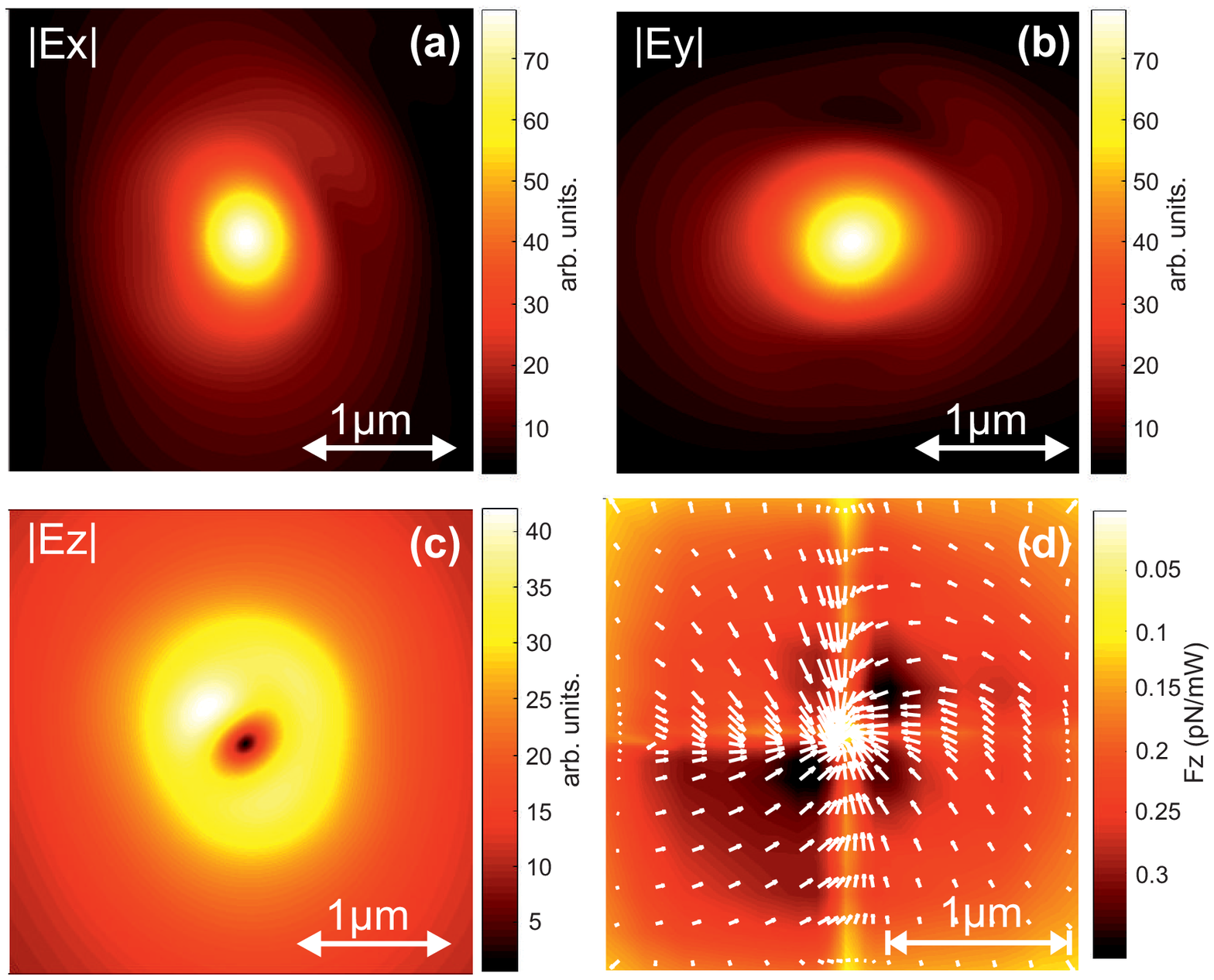}
\caption{ {\textbf{Electric field and force distributions} at $\lambda=1080$~nm. \textbf{(a-c) Maps of the electric field component amplitudes} in a transverse plane ($xy$ plane) located at $D=305$~nm from the OSTT apex. \textbf{(d) Distribution of the  force components} in the same plane : vertical component $F_z$ in color level while white arrows correspond to the transverse force.}}
\label{Fxy}\rule{\textwidth}{0.4pt}
\end{figure}
\begin{figure}
\centering
\includegraphics[width=16cm]{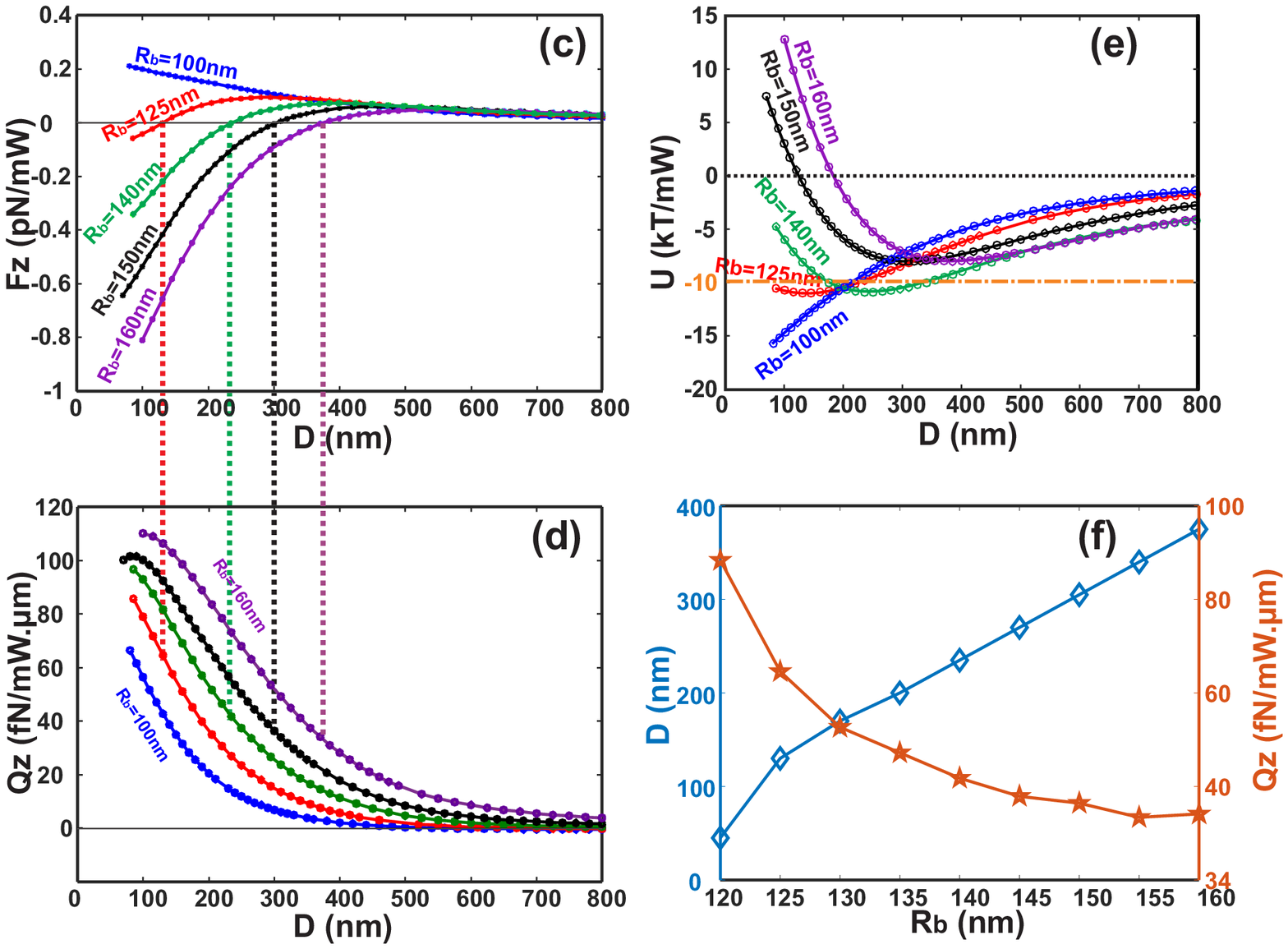}
\caption{\textbf{Optical force, optical torque, and potential variations. (a) Vertical force variations} exerted on a latex NP as a function of $D$ for different values of the NP radius $R_b$. In all cases, the refractive optical index of the latter is fixed to $n_b $~=~1.5. \textbf{(b) Magnitude of the z-component of the torque} as a function of $D$ for the same $R_b$ values as in (a). Vertical dashed lines denote stable trapping that occurs for all considered values of $R_b$ except $R_b $~=~100~nm. \textbf{(c) Potential derived from the optical force} as a function of $D$ for the same $R_b$ values as in (a) and (b). Notice that a value of 10$kT$ is generally assumed for a stable trapping to compensate the Brownian motion. \textbf{(d) Trapping distance and torque} as a function of $R_b$. In cases where trapping occurs, the torque is never equal to zero.}
\label{torqueforceRb}\rule{\textwidth}{0.4pt}
\end{figure}

Figure \ref{Rb150}b shows the vertical components of both the optical force $F_z$ in solid blue line, and of the torque $Q_z$ in dashed red line as a function of $D$ for $\lambda $~=~1080~nm. As predicted, the force cancels out for $D $~=~305~nm \textcolor{black}{(contactless trapping as obtained within a diabolo nano-antenna in ref. \cite{hameed:sr17})} but not the torque whose value is equal $Q_z $~=~35.3~fN/mW.$\mu$m. This positive value is consistent with the rotation direction of the electric field associated with the circularly polarized wave emitted by the OSTT. Figure \ref{Rb150}c corresponds to this trapping-spinning state (point A of Fig. \ref{Rb150}a) and shows the electric intensity distributions in three perpendicular planes at the operation wavelength for a linearly polarized incident guided mode along $\alpha $~=~45$^\circ$ (the angle $\alpha$ is shown in Fig. \ref{schemasonde}a). As expected, the mode inside the aperture is excited in both $xz$ and $yz$ planes even though  {its dimension is} not the same in both planes. \\

{Note here that the numerical results seem to demonstrate a momentum transfer that induces the spinning of the particle. This is generally forbidden when considering a spherical purely dielectric particle made in isotropic medium and illuminated by a plane wave even if the latter is circularly polarized. \textcolor{black}{As explained in the introduction, the reason for this transfer lies, in our case, in the fact that the FDTD mesh inevitably leads to a symmetry breaking for any three-dimensional object which is estimated, for a given direction, to be equal to the spatial step of discretization along this direction. Thus, in our case, the spatial step being $5~nm$, this makes that the NP could have an elliptical shape of major axis $R+5~nm$ and a minor axis of $R -5~nm$. In addition,} the distribution of the electric field in the transverse plane (perpendicular to the axis of the probe) presents an asymmetry induced by the elliptical geometric shape of the aperture itself. To be convinced, we have plotted on Fig. \ref{Fxy}a-c the distribution of the three components of the electric field amplitude in such a plane. As can be seen, these distributions do not show central symmetry especially the normal component ($E_z$) which looks like a distorted torus. The transverse components ($E_x$ and $E_y$) lead to a perfectly circular polarization in the central region where the $E_z$ component becomes zero. 
In order to demonstrate that the trapping exists at the center (probe axis at $x=y=0$) in the transverse plane corresponding to $Fz=0$, we considered a particle sweeping this plane and we calculated the optical force on it for each position at different values of the wavelength. Figure \ref{Fxy}d shows the vertical force $F_z$ (in color level) and the transverse force component (white arrows) in the plane corresponding to $D=305~nm$ at $\lambda=1080nm$ for a particle radius of $R_b=150~nm$ moving between $-1.24\mu m~$ and $1.24\mu m$ along both the $x$ and $y$ directions. It can clearly be seen that the particle is pushed toward the axis $x=y=0$ where $F_z=0$. The movie \textbf{"Visualization3.avi"} shows the evolution of this transverse force as a function of the wavelength varying from $600nm$ to $1200nm$. By looking closely to this video, we can see that the trapping only occurs around $\lambda=1080nm$ for which the aperture behaves as a quater-wave plate.} \\

Figures \ref{torqueforceRb}a and b present the z-component of the force $F_z$ and the torque $Q_z$ respectively for different values of the NP radius $R_b$.  Qualitatively, the force amount is comparable to that found in Ref. \cite{saleh:nl12} with coaxial aperture {(around 50 pN/100mW for a small bead of 15 nm diameter}) and it is at least twice as great as that obtained theoretically and experimentally with the bowtie nano-aperture antenna of Ref. \cite{eleter:oex14a}. The origin of this high efficiency is linked to the excitation of a frozen (small group velocity) guided mode inside the coaxial aperture resulting in significant energy funneling (large radiation pressure) together with high electric field confinement (gradient force). The torque seems to present an almost exponential decreasing with $D$ as shown in Fig. \ref{torqueforceRb}d. \\

To quantify the trapping strength, we plot on Fig. \ref{torqueforceRb}c the variations of the associated scalar potential (defined by $\vec{F}=-\vec\nabla(U)$) from which the optical force derives. One can clearly see that when trapping occurs ($R_b\in$[125;160]nm), a power source less than 1.5 mW leads to highly stable trapping for which the potential well is deeper than -10$kT$ {\cite{ashkin:ol86}} (horizontal orange dashed-dotted line). The evolution of the stable trapping distance as a function of the NP radius $R_b$ is given in Fig. \ref{torqueforceRb}d (blue curve) in addition to the corresponding z-component of the torque for each NP radius. As expected, when the radius increases, the stable trapping position increases, and the torque decreases. Nevertheless, the behavior of these two variations is different: it is a fairly linear function of $D$ while it is almost exponential for the torque.  
As mentioned earlier, a conventional polarization controller can be used to tune the rotation frequency or even invert its sign (and even cancel it if desired). To demonstrate this effect, we have performed FDTD simulations by modifying the polarization direction of the guided mode defined by the angle $\alpha$ (see inset of Fig. \ref{schemasonde}a) as it can be done using such a controller. In these simulations, the NP has an optical index of $n_b $~=~1.5 and we set its radius to $R_b $~=~150~nm while the distance to the OSTT is $D $~=~305~nm corresponding to the stable distance trapping previously calculated (see Fig. \ref{torqueforceRb}b). Figure \ref{polar}a shows a schematic view of the studied configuration. The variations of the z-components of the force and torque are presented in Fig. \ref{polar}b. As expected, the force is quite constant and its value is very small (smaller than 0.01~pN/mW), so that the trapping distance is almost independent of the direction of polarization. The torque, on the other hand, undergoes a change in amplitude and a sign inversion when the incident polarization aligns with the axes of the ellipse (transmitted beam is then linearly polarized). As shown in figure \ref{polar}b, the behavior is not symmetrical as in the ideal case of a quarter-wave plate where the torque \textcolor{black}{should vary} as $sin$ 2$\alpha$. This is due to the non-perfect geometry of the probe (aperture not perfectly aligned with the fiber axis \& a symmetry of revolution which is broken by the mesh used in the FDTD \cite{hoblos:ol20}) which leads to a cancellation of the angular momentum for an angle $\alpha$~=~82$^\circ$ instead of 90$^\circ$.    
\begin{figure}
\centering
\includegraphics[width=16cm]{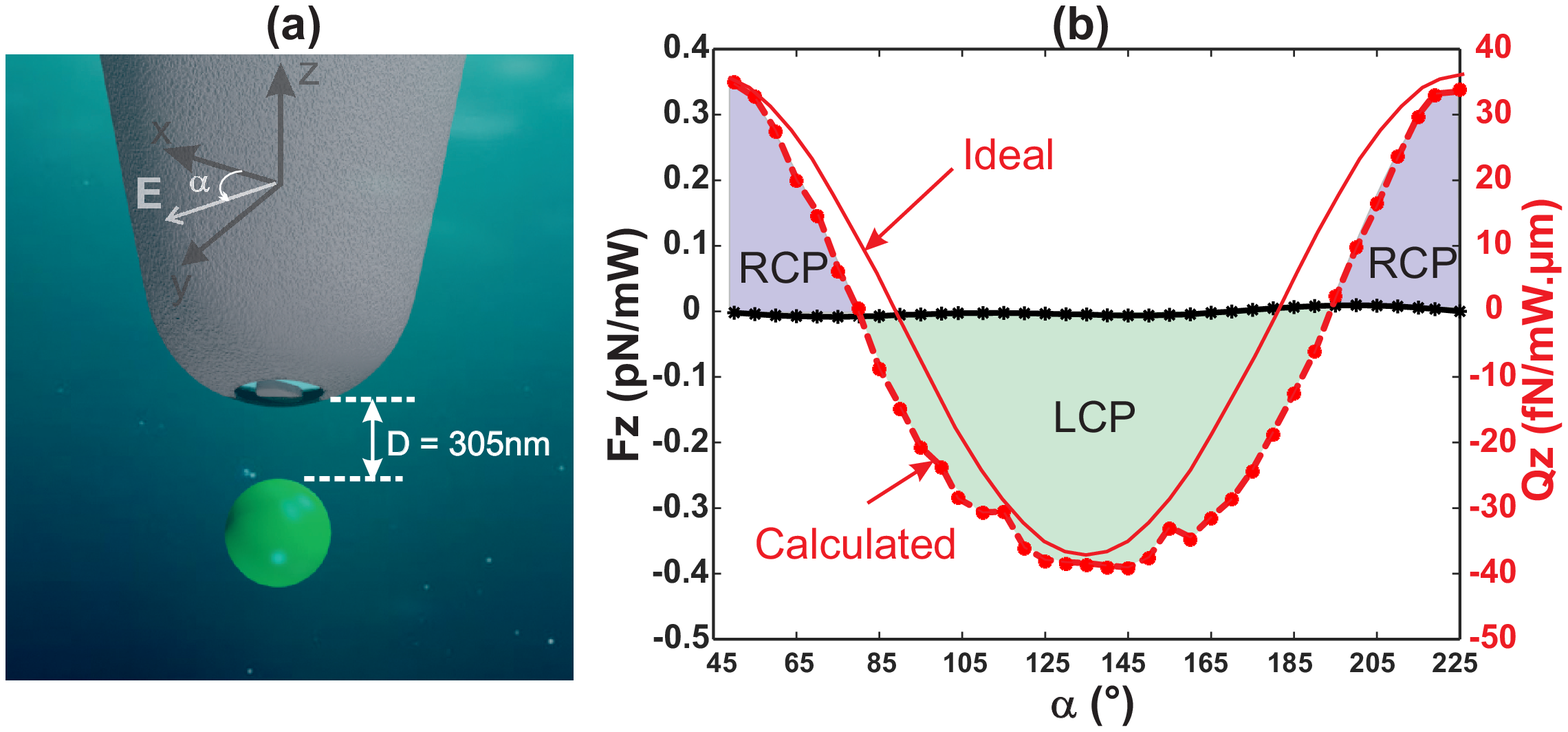}
\caption{\textbf{Trapping and spinning properties} as a function of the polarization direction of the guided mode inside the fiber given by the angle $\alpha$. \textbf{(a) Schematic of the studied configuration} corresponding to a stable trapping of a NP radius $R_b $~=~150~nm at a distance $D $~=~305~nm. \textbf{(b) Variations of the z-components of the force (in black solid line) and the torque (red dashed line)} as a function of the guide mode polarization direction $\alpha$. The solid red line corresponds to the torque variations within a perfect quart-wave plate.} 
\label{polar}
\end{figure}
Figure \ref{torqueforceRb}c shows that an injected power of 5 mW allows guaranteeing the stable trapping criterion of $U>10KT$ {\cite{ashkin:ol86}} for NP radius $R_b$ between 125~nm and 160~nm. The optical torques (median value of 25~fN$\cdot\mu $m for only 1~$mW$ injected power) are then as big as the one measured in Ref. \cite{monteiro:pra18} ($Q_z$~=~28$\pm$13~$fN\cdot\mu$m) using a bulky configuration. This suggests an excellent spinning efficiency even if the NP is in water. In fact, the relation \textcolor{black}{that corresponds to} the steady-state rotation motion (torques equilibrium) in the case of a pure dielectric sphere (without absorption) is given by \cite{reimann:prl18,corson:pre17}:
\begin{equation}
\tau^{sphere}_{drag}=-Q_{z}=-\frac{4\pi^2\mu R_b^4\nu_r}{1.497\Lambda}
\end{equation}
Where $\tau^{sphere}_{drag}$ is the viscous torque for a $R_b$-radius hard-sphere rotating at frequency $\nu_r$ in a liquid of dynamic viscosity $\mu$ with a molecular mean free path of $\Lambda$. If we consider a $R_b$~=~125~nm-radius NP immersed in water ($\mu=\mu_{water}$~=~8.9$\times$ 10$^{\text{-4}}$~Pa.s, $\Lambda=\Lambda_{\text{water}}$~=~2.5$\AA$) trapped within an injected power of 1~mW, the spinning frequency of the rotational steady-state is evaluated to be $\nu_r^{\text{water}}$~=~28.2~Hz. This relatively small value of the rotation frequency can be greatly enhanced when NP is trapped in air (up to $\nu_r^{\text{air}}$~=~3.34~MHz for $\mu_{\text{air}}$~=~1.8$\times$ 10$^{\text{-5}}$ and $\Lambda_{\text{air}}$~=~0.6~$\mu$m) and obviously higher values \textcolor{black}{could be obtained} in vacuum.
\section{Conclusion}
In summary, the proposed OSTT offers a unique fibered/miniaturized design of an optical tweezers able to act as a quarter-wave nano-antenna inducing a contactless trapping together with a spinning of a NP. The geometric parameters of the OSTT can be adapted to operate over a wide spectral range as well as according to the physico-chemical nature of the NP and/or of the host medium. The metal coating can also be treated to inhibit any chemical reaction (oxidation, sulfurization,...). Nevertheless, one should be aware of the fact that the presence of metal could induce harmful effects (heating then destruction of the probe) in the case of high injected power. This is why the coaxial aperture remains an excellent compromise due to its high transmission efficiency allowing to use it as an OSTT with low optical power in many \textcolor{black}{optomechanics applications (optical pressure control, sorting of chiral and no chiral particles, translational and rotational nano-positioning,...)} that require relatively low light powers.
\section*{acknowledgments}
Computations have been performed on the supercomputer facilities of the "M\'esocentre de calcul de Franche-Comt\'e". This work has been partially supported by the EIPHI Graduate School (contract ANR-17-EURE-0002). Mrs. Zaghdoudi thanks the University Mouloud Memmeri of Tizi-Ouzou in Algeria for financing her doctoral internship at FEMTO-ST Institute of Besan\c con (France). 
\section*{Appendix A: Simulation method}
We use home-made FDTD codes that were adapted to the studied geometry. The FDTD is based on the solving of Maxwell equations through a temporal iterative scheme. The space is discretized into small cells of dimension as small as $\lambda/30$ and the electromagnetic field is calculated for each cell through a centered finite-difference schema \cite{taflove:book05}. Due to the temporal character of the FDTD, dispersive materials need to be modeled through an analytical model giving their permittivity as a function of the wavelength. The model is then integrated into the FDTD algorithm using the auxiliary differential equation method or the recursive convolution method. In the case of the grating of Fig. \ref{grating}, periodic boundary conditions are applied in the $x$ and $y$ directions while absorbing (Perfectly Matched Layer technique) ones are used in the $z-$direction to cancel parasitical reflections on the limits of the simulation window. For the OSTT, a Full 3D-FDTD code is used where the absorbing boundary conditions are applied in the three directions of the space due to the finite character of the configuration. In this case, the simulations are much heavier and require a calculation time of about 3 days per NP position. In both codes, we used a non-uniform mesh allowing us to describe the geometry of the structure (cell size) as well as possible. The size of the cell varies from $5~nm$ near the probe apex and the NP to $20~nm$ elsewhere. The force and torque calculations are done using Matlab software through a code that allows implementing Eqs. 1. The integration is made over a cube which encompasses the NP and having all its faces in the same medium (here water or air) in order to fulfill the condition of conservation of the force to be able to use divergence theorem which allowed us to transform the volume integral into a surface integral. In all simulations, the side of the cube was set to the diameter of the NP increased by 6 FDTD cells (3 cells on each side).
\begin{figure}
\centering
\includegraphics[width=13cm]{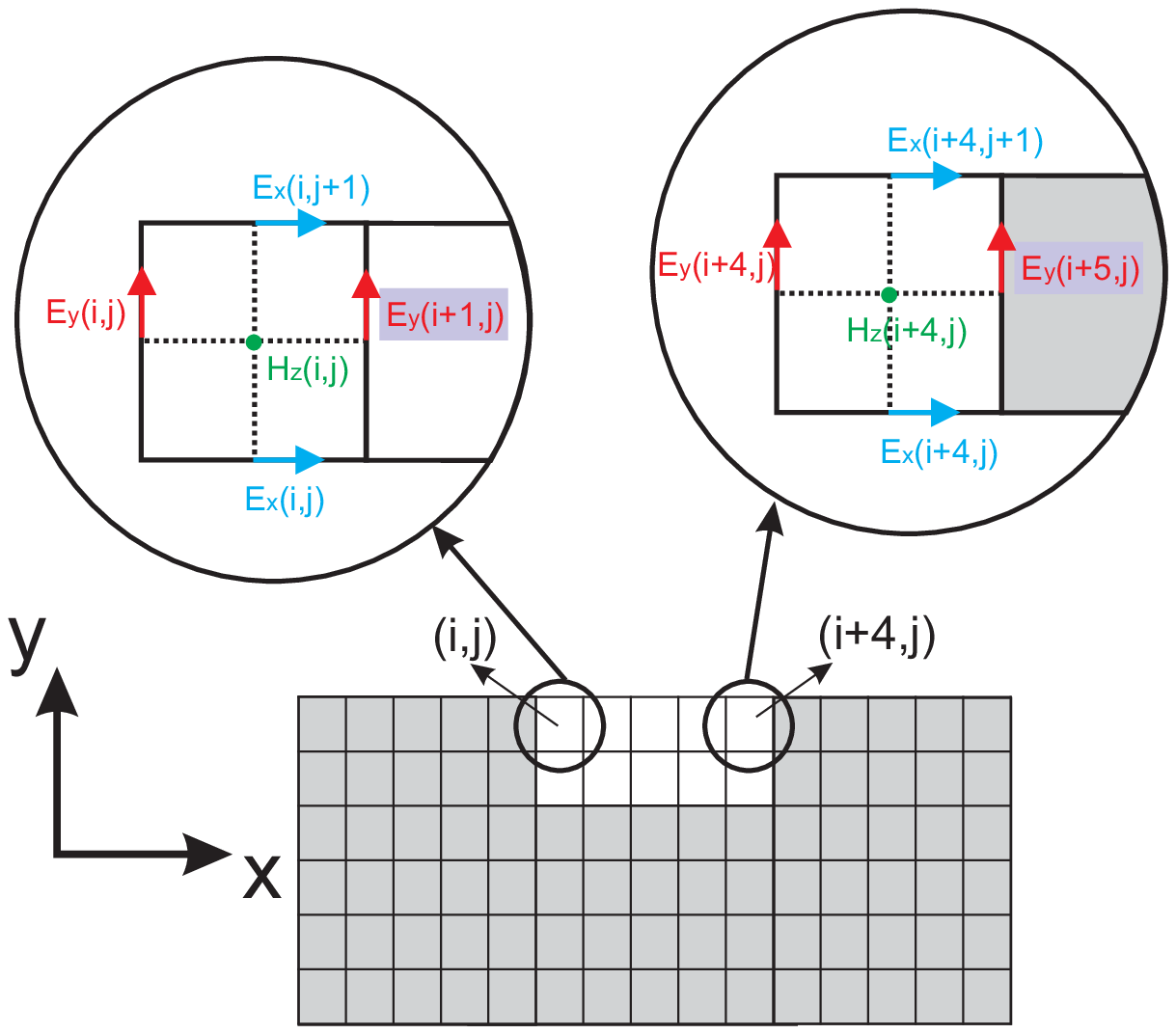}
\caption{\textbf{FDTD intrinsic symmetry breaking.} The scheme of an FDTD mesh applied to a geometrically symmetric structure (gray and white materials). The Yee schema used to calculate the electromagnetic components is at the origin of a numerically induced dissymmetry. } 
\label{SPM_revealing}
\end{figure}
On the other hand, to illustrate the FDTD intrinsic symmetry breaking mentioned in the introduction, we present on figure \ref{SPM_revealing} a meshing in 2D of a vertical-axis symmetrical structure consisting in two different materials (gray and white). The Yee schema used to calculate the electromagnetic field of a cell is also emphasized for two specific cells (show the big circles). 
The expression of the corresponding magnetic fields is given by:\\
\begin{eqnarray}{ll}
H_z(i,j,t)&=& H_z(i,j,t-\Delta t) \notag \\
&+&\cfrac{1}{\mu}\bigg\lbrace\cfrac{E_x(i,j+1,t-\Delta t/2)-E_x(i,j,t-\Delta t/2)}{\Delta y} \\
&-&\cfrac{E_y(i+1,j,t-\Delta t/2)-E_y(i,j,t-\Delta t/2)}{\Delta x} \bigg\rbrace \notag \\ 
H_z(i+4,j,t)&=&H_z(i+4,j,t-\Delta t)\notag \\
&+&\cfrac{1}{\mu}\bigg\lbrace\cfrac{E_x(i+4,j+1,t-\Delta t/2)-E_x(i+4,j,t-\Delta t/2)}{\Delta y}\label{hupdate}\\
&-&\cfrac{\textcolor{red}{E_y(i+5,j,t-\Delta t/2)}-E_y(i+4,j,t-\Delta t/2)}{\Delta x}\bigg\rbrace \notag 
\end{eqnarray}

When illuminating the structure by a plane wave propagating along the $y$ direction (see Fig. \ref{SPM_revealing}), the two magnetic field ($H_z$) of these two cells $[(i,j)$ and $(i+4,j)]$ should be equal. Unfortunately, the Yee schema does not allow to fulfill this condition due to the fact that the electric field component $E_y(i+5,j)$, highlighted in red in Eq. \ref{hupdate} and used to calculate the corresponding magnetic field of the cell $(i+4,j)$, is relative to the gray material while both $E_y$ components of the cell $(i,j)$ correspond to the white material. This break of symmetry exists even if the whole configuration is symmetrical. Nevertheless, the smaller the size of the FDTD cell, the smaller the symmetry breaking \cite{hoblos:ol20}.
\section*{Appendix B: Optimization of the geometry}
 {The approach used is that described in Ref. \cite{dahdah:ieee12}}. We start by considering a 2D square lattice grating of elliptical coaxial apertures (see Fig. \ref{grating}a) deposited on a glass substrate and we adapt the metal thickness by taking water as superstrate. In fact, for trapping applications, NPs are generally immersed in a liquid to counteract their weight by the buoyancy forces. Home-made FDTD codes were then used to simulate the light transmission through this grating. A schema of the modeled structure is presented in Fig. \ref{grating}a; the geometrical parameters and notations are shown on the sketch of the top-view of Fig. \ref{grating}b and their values are given in the caption. The glass substrate refractive index is fixed to $n_s$~=~1.458 and the dispersion of the silver film is described through a Drude-Critical Points (DCP) model \cite{Hamidi2011} which has been adapted to the studied spectral range ($\lambda\in$[600~nm,1400~nm]). 
Figure \ref{grating}c shows a diagram giving the transmission spectrum as a function of the metal thickness $t$. Similarly to Ref. \cite{baida:prb11}, we superimposed dotted green lines corresponding to a phase difference of $\Delta\phi=\pi/$2 between the two transmitted transverse components of the electric field; essential condition to have a quarter wave plate. Moreover, the cyan lines correspond to another condition for which the transmission coefficients of \textcolor{black}{these two components} are identical. This fulfills the requirement on the transmitted amplitudes to obtain a circular polarization instead of an elliptical {one}. All intersections of the green {lines} with the cyan lines give solutions that fulfill a quarter-wave plate behavior. The last step is then to choose the intersection point corresponding to the higher value of the transmission coefficient. From Fig. \ref{grating}c, the more adapted thickness is obtained for $t=95~nm$ and the operation wavelength value is then around $\lambda=1093.5~nm$ for which transmission spectra and phase difference are shown on Fig. \ref{grating}d. Note that the geometry has been adapted to obtain an operating wavelength value corresponding to low light absorption by water \cite{eleter:oex14a,hameed:ieee14}.
\textcolor{black}{Considering a single elliptical coaxial aperture etched at the apex of a metal-coated SNOM probe, the transmission properties, in terms of polarization, deviate slightly from those of a periodic structure (same aperture grating), especially because the lighting conditions are not the same (fiber mode in the case of the probe instead of a plane wave in the case of the grating). For this reason, the geometric parameters have been adapted leading to an operating wavelength around 1080 nm. This was achieved through additional\textcolor{black}{ full 3D-}FDTD simulations with the same aperture geometry leading to a slightly different value of $h = 125 nm$ instead of $95 nm$ as presented in Fig. \ref{schemasonde}b.}
\begin{figure}
\centering
\includegraphics[width=14cm]{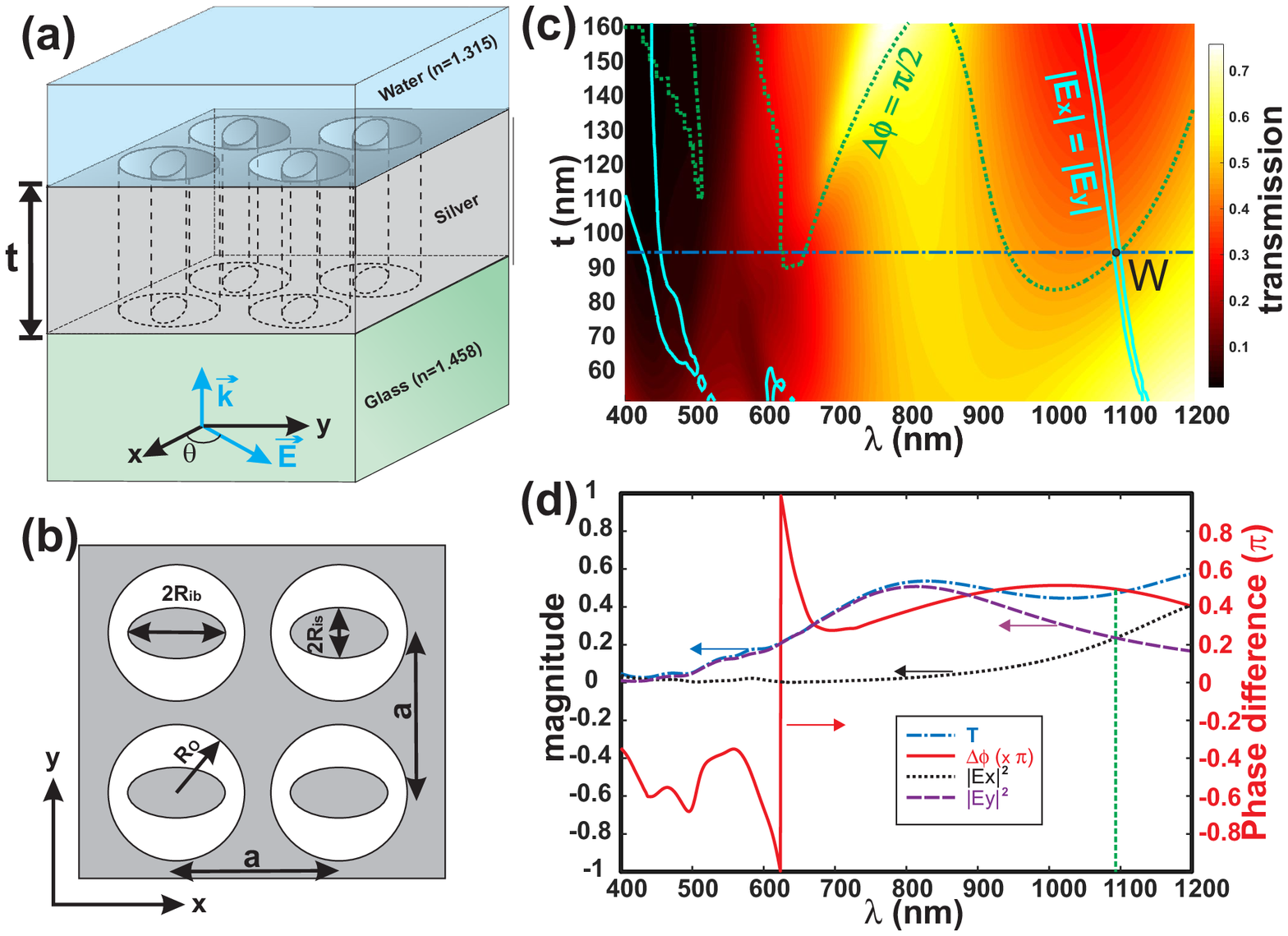}
\caption{\textbf{Polarization conversion through a metallic grating of \textcolor{black}{coaxial} elliptical apertures. (a) Schema of the studied silver grating} deposited on a glass substrate and immersed in water (superstrate). \textbf{(b) Top view sketch of the grating} giving its geometrical parameters ($a$~=~300~nm is the period, $R_{ib}$~=~80~nm is the major-axis of the inner ellipse, $R_{is}$~=~50~nm is its minor-axis while $R_o$~=~120~nm is the {radius of the outer circle of the aperture}).  \textbf{(c) Transmission coefficient spectra} as a function of the metal layer thickness \textcolor{black}{$h$}. The cyan contour lines are calculated for equivalent intensities of the two $x-$ and $y-$ components of the transmitted electric field while the dotted green lines denote couples (\textcolor{black}{$h$},$\lambda$) for which the phase difference is equal to $\pi/2$. The {blue dashed-dotted horizontal line passing} through the operation point $W$ corresponds to a metal thickness \textcolor{black}{$h$}~=~95~nm. \textbf{(d) Transmission spectra} of the two electric field components, $E_x$ in black dotted line and $E_y$ in purple dashed line, in addition to the total transmission $T$ in blue dotted-dashed line for a metal thickness of \textcolor{black}{$h$}~=~95~nm corresponding to the horizontal blue dotted-dashed line on (c). The phase difference $\Delta\phi$ is also plotted in red solid line. As expected, the intersection between the black and the purple curves corresponds exactly to a phase difference of $\pi$/2 as indicated by the vertical green dashed line.}
\label{grating}
\end{figure}

\end{document}